\documentclass[12pt]{article}
\usepackage{amsmath}
\usepackage{graphicx,psfrag,epsf} 
\usepackage{enumerate}
\usepackage{url} %

\usepackage[utf8]{inputenc} %
\usepackage[T1]{fontenc}    %
\usepackage{hyperref}       %
\usepackage{booktabs}       %
\usepackage{amsfonts}       %
\usepackage{nicefrac}       %
\usepackage{microtype}      %
\usepackage{times}
\usepackage{amsmath}
\usepackage{subfigure} 
\usepackage{float}
\usepackage{xcolor}
\usepackage{algorithm}
\usepackage{algorithmic}
\usepackage{hyperref}
\usepackage{bm}
\usepackage{bbm}
\usepackage{natbib}
\usepackage{enumitem}
\usepackage[toc,page]{appendix}
\usepackage[font=small,labelfont=bf]{caption}
\RequirePackage[normalem]{ulem} 

\newcommand{\blind}{1}

\ifx\BlackBox\undefined
\newcommand{\BlackBox}{\rule{1.5ex}{1.5ex}}  %
\fi

\ifx\QED\undefined
\def\QED{~\rule[-1pt]{5pt}{5pt}\par\medskip}
\fi

\ifx\proof\undefined
\newenvironment{proof}{\par\noindent{\bf Proof\ }}{\hfill\BlackBox\\[2mm]}
\fi

\ifx\theorem\undefined
\newtheorem{theorem}{Theorem}
\fi
\ifx\example\undefined
\newtheorem{example}[theorem]{Example}
\fi
\ifx\property\undefined
\newtheorem{property}{Property}
\fi
\ifx\lemma\undefined
\newtheorem{lemma}[theorem]{Lemma}
\fi
\ifx\proposition\undefined
\newtheorem{proposition}[theorem]{Proposition}
\fi
\ifx\remark\undefined
\newtheorem{remark}[theorem]{Remark}
\fi
\ifx\corollary\undefined
\newtheorem{corollary}[theorem]{Corollary}
\fi
\ifx\definition\undefined
\newtheorem{definition}[theorem]{Definition}
\fi
\ifx\conjecture\undefined
\newtheorem{conjecture}[theorem]{Conjecture}
\fi
\ifx\fact\undefined
\newtheorem{fact}[theorem]{Fact}
\fi
\ifx\claim\undefined
\newtheorem{claim}[theorem]{Claim}
\fi
\ifx\assumption\undefined

\fi
\numberwithin{equation}{section}
\numberwithin{theorem}{section}
\usepackage{multirow}

\newcommand{\bx}{\mathbf{x}}
\newcommand{\bR}{\mathbb{R}}

\newcommand{\bbeta}{\bm{\beta}}
\newcommand{\balpha}{\bm{\alpha}}
\newcommand{\bgamma}{\bm{\gamma}}

\newcommand{\bd}{\mathbf{d}}

\newcommand{\bfeta}{\bm{\eta}}
\newcommand{\cN}{\mathcal{N}}
\newcommand{\mS}{\mathcal{S}}
\newcommand{\argmax}{\mbox{argmax}}

\newtheorem{assumption}{Assumption}

\begin{document}

\if1\blind
{
\title{\bf Model-based Sparse Coding beyond Gaussian Independent Model
}

\author
{
    Xin Xing\thanks{ Department of Statistics, Virginia Tech},    $\quad$
   Rui Xie \thanks{Statistics and Data Science, University of Central Florida},   $\quad$
   Wenxuan Zhong\thanks{Department of Statistics, University of Georgia.}
}
\maketitle
} \fi

\begin{abstract}
Sparse coding aims to model data vectors as sparse linear combinations of basis elements, but a majority of related studies are restricted to continuous data without spatial or temporal structure. A new model-based sparse coding (MSC) method is proposed to provide an effective and flexible framework for learning features from different data types: continuous, discrete, or categorical, and modeling different types of correlations: spatial or temporal. The specification of the sparsity level and how to adapt the estimation method to large-scale studies are also addressed.  A fast EM algorithm is proposed for estimation, and its superior performance is demonstrated in simulation and multiple real applications such as image denoising, brain connectivity study, and spatial transcriptomic imaging.  
\end{abstract}

\vfill
\section{Introduction}\label{sec:intro}
Sparse coding aims at decomposing an $m$-dimensional random vector as a linear combination of $K$ interpretable vectors, a collection of which is also referred to as a dictionary. Each vector in a dictionary is referred to as an atom \cite{aharon2006img,mairal2009online}. Compared to the wavelet or kernel estimation methods which use predefined basis functions~\cite{mallat1989theory}, sparse coding is more interpretable and flexible, which leads to state-of-the-art discoveries in numerous scientific fields such as neuroscience, genomics, artificial intelligence and astronomics~\cite{lee2011data,carin2012high,zhang2010discriminative,bi2014sar}. 

Sparse coding is a special case of matrix factorization. Sparse coding techniques consider
a finite series of input signals $\bx_i\in\mathbb{R}^m$, $i=1,\dots, n$, and optimize the empirical loss function 
\begin{equation}\label{eq:classical}
L(\bbeta_1, \cdots, \bbeta_n, D)=\frac{1}{n}\sum_{i=1}^n(\frac{1}{2}||\bx_i - D\bbeta_i||^2 + \lambda||\bbeta_i||_1),
\end{equation}
where $\|\cdot\|$ is the Euclidean norm, $D\in \mathbb{R}^{m\times K}$ is the dictionary, $\bbeta_i\in\mathbb{R}^K$ are the corresponding coefficients for the $i$th signal, and $\lambda$ is a trade-off between the sparsity of $\bbeta_i$s and goodness of fit. 

\subsection{Related work}
Sparse coding algorithms have been studied over decades. The majority of existing algorithms aim at solving the optimization problem by estimating the dictionary and coefficients alternatively. 
A few popular methods along this line of thinking are K-SVD \cite{aharon2006img}, online dictionary learning \cite{mairal2009online}, and recursive least squares \cite{skretting2010recursive}, all of which first estimate sparse coefficients by either the matching pursuit algorithm \cite{cotter1999forward} or the orthogonal matching pursuit \cite{pati1993orthogonal} algorithm and then update the dictionary through the coordinate descent approach. These algorithms have shown state-of-the-art performance in imaging denoising, inpainting, and super-resolution.  However, in many applications such as spatial transcriptomics and neuroimaging, data are often discrete or have correlated noise. Considering that these data structures may significantly increase the model interpretability and accuracy of the learned features, many existing algorithms have some restrictions.  For example, one must assume that the noise is independent with fixed variance, that input signals are continuous, and that the sparsity level is predefined or determined by cross-validation. Classically, spatial data analysis such as that done in \cite{gotway1997generalized, schabenberger2017statistical} are widely used in generalized linear models (GLMs). GLMs highly depend on fixed bases, which are usually unavailable in many applications. There is a lack of methods modeling spatial information in a sparse coding model which can learn data-adaptive bases. 
To mitigate such limitations, we propose a model-based sparse coding (MSC) algorithm, in which both the continuous and discrete input with various noise distributions can be handled by different probabilistic models. We use the Bayesian information criterion (BIC) for determining the sparsity of the model.  

Our work is also related to the probabilistic models which have been widely used in solving various matrix factorization problems such as principal component analysis (PCA)  \cite{tipping1999probabilistic,guan2009sparse}, non-negative matrix factorization (NMF) \cite{shashanka2008probabilistic, fevotte2009nonnegative} and singular value decomposition (SVD) \cite{mnih2008probabilistic}.
However, unlike most probabilistic models with only a small number of components, MSC has a large number of potential components, each of which is a possible combination of atoms. For example, if we assume the $l_0$ norm of the coefficients are at most $d$, there are $\sum_{l=1}^d\binom{K}{l}$ possible combinations of atoms to represent the input signal. The total number of combinations increases exponentially with $d$, which makes the estimation computationally infeasible. To break the computational bottleneck, we propose a fast EM algorithm that reduces the computational cost from an exponential to a linear order of $d$ by sequentially increasing $d$. For a fixed $d$, we use a rejection-control strategy to reduce the computational cost further. 

\subsection{Our contribution}
The MSC approach is a rich and flexible framework under which many data types and various noise distributions can be handled easily. 
Despite the continuous data, discrete data are often seen in many applications. Using  discrete distributions such as binomial or Poisson will better fit the data and provide more interpretable results.
For example, in the spatial transcriptomic data, input signals are the counts of mapped reads, and we applied MSC with a mixture of Poisson distributions. Poisson distributions are often dictated by the nature of counts data in biological studies and will avoid negative values in the outcome, which are difficult to interpret in real applications. 

Moreover, spatial or temporal patterns are usually of interest and provide rich information in real applications. Proper modeling of this pattern may significantly increase the model interpretability and the accuracy of learned features.
For example, in spatial transcriptomic data, the spatial pattern of proteins or messenger RNAs (mRNAs) plays a vital role in biomedical research and diagnostics. By considering this spatial information, our proposed MSC model provides a clear separation of invasive and noninvasive cancer areas, which is a challenging task in cancer diagnosis. In another real example presented herein, we consider the temporal correlation in functional magnetic resonance (fMRI) imaging data. By incorporating these correlations using a data-adaptive covariance structure, MSC learns meaningful brain network structures that are verified through a comparison with existing templates. 

To make MSC computationally feasible for massive datasets, we also propose a fast EM algorithm that can reduce the computational cost to the linear order of $d$. The superior performance of MSC in image processing and biomedical studies is established in our simulated and real examples.

\subsection{Outline of the paper}
The rest of the article is organized as follows. Section 2 will formally propose the probabilistic sparse coding model. An algorithm based on the model and discussion of the opportunities and challenges raised by the algorithm will be discussed in Section 3.  Simulations and applications will be collected in Section 4. 

\section{Model Set-up}

\def \bgamma{\boldsymbol{\gamma}}
Assume the $i$th signal, $\bx_i=(x_{i1},\dots,x_{im})^\top\in\mathbb{R}^m$, follows a mixture distribution, of which each component is distributed from $f(\bx_i|\theta_{ij})$ for $j=1,\cdots, J$. The fractions of each component are $(\pi_1, \cdots, \pi_J)$. Notation-wise, we  write 
\begin{equation}\label{mix}
\bx_i\sim \pi_1f(\bx_i|\theta_{i1})+\cdots+\pi_Jf(\bx_i|\theta_{iJ}).
\end{equation}

A binary membership labeling variable $z_{ij}$ for observation $i$ can be introduced such that $\bx_i|z_{ij}=1\sim f(\bx_i|\theta_{ij})$. For example, if each component in (\ref{mix}) follows a multivariate Gaussian distribution with mean $D\balpha_{ij}$ and covariance $\sigma_i^2 I$,  $\theta_{ij}$ includes  $\{D, \balpha_{ij}, \sigma_i\}$, where $D\in\mathbb{R}^{m\times K}$ is the dictionary and $\balpha_{ij}\in \mathbb{R}^K$ is the corresponding coefficient.  Given the latent membership $z_{ij}=1$, we have the mean $\mathbb{E} [x_i\mid z_{ij}=1]= D\bm{\alpha}_{ij}$, corresponding to the representation of $x_i \approx D\beta_i$ in classical dictionary model (\ref{eq:classical}).

The model (\ref{mix}) is not estimable without constraints on $\balpha_{ij}$, as there are more parameters than observations. However, when $\balpha_{ij}$ satisfies some sparsity constraints, the number of parameters will significantly reduce, and the model (\ref{mix}) will be estimable. The proposed model is general, where many popular models can be considered as special cases. If we assume $||\balpha_{ij}||_0=1$, i.e., each component is only related to one atom, the model (\ref{mix}) is equivalent to the model-based clustering \cite{banfield1993model}. If we further assume $||\balpha_{ij}||_0\le d$,  the $i$th signal follows a mixture distribution with $J=\sum_{l=1}^d\binom{K}{l}$ components, each of which is only related to at most $d$ atoms.  
Thus, the model (\ref{mix}) can be cast as a probabilistic version of the sparse coding model. 
To obtain the optimal sparse model, we only need to estimate $\pi_1$ to $\pi_J$ to see which one is larger. 

In the following subsections, we will first discuss the model-based sparse coding when $f(\bx_i|\theta_{ij})$ is a Gaussian density with covariance matrix $\sigma_i^2I$. Then we will generalize the Gaussian mixture model to incorporate spatial correlations. Finally, we will generalize the Gaussian mixture to the distributions from the exponential family to model discrete data.  In practice, we first choose between the exponential family or the Gaussian model based on the input data type. If the input data is discrete count, we prefer to choose the exponential family $d$-sparse model. Otherwise, for continuous data, we choose Gaussian $d$-sparse model. Based on whether spatial or temporal information exists in the data, we could choose between the spatial and simple $d$-sparse models.

\subsection{Simple $d$-sparse Gaussian MSC}
We first consider the simple case when the input signals $\bx_i$ are continuous and $x_{i1},\dots,x_{im}$ are mutually independent. 
Mathematically, we can formulate the simple $d$-sparse Gaussian MSC as 
\begin{multline}\label{model1}
\bx_i\sim \pi_1\cN(\bx_i|D(\balpha_{i1}\circ\bgamma_1), \sigma_i^2I)+\cdots 
+\pi_J\cN(\bx_i|D(\balpha_{iJ}\circ\bgamma_J), \sigma_i^2I),
\end{multline}
where $\balpha_{ij}\in\bR^{K}$, $\bgamma_j=(\gamma_{j1}, \cdots, \gamma_{jK})^\top$ is a $K$-dimensional binary vector that controls which atoms are selected in the $j$th component.  Here, we use $\circ$ to denote the Hadamard product. To achieve $d$-sparsity, we require $\sum_{l=1}^K\gamma_{jl}\le d$. It is easy to see that model (\ref{model1}) includes all possible $d$-sparse combinations of atoms, where each combination is a specific component in (\ref{model1}). Notice that when the number of atoms, $K$, is large, the number of all possible $d$-sparse combinations is a large number. In real applications, some of the observed $n$ signals likely share the same combination of atoms. We introduce the following assumption in which we assume that the number of $d$-sparse combinations in (\ref{model1}) is small.
\begin{assumption}\label{aa1} $\gamma_j$ is a K-dimensional binary vector defined in (\ref{model1}). We assume that
\begin{equation}
\lim_{n\to \infty} |\{\gamma_j\}|/n \to 0 , 
\end{equation}
where $|\cdot|$ denotes the cardinality of the set. 
\end{assumption}

We propose a new fast EM algorithm that sequentially searches the possible $d$-sparse combinations based on the $(d-1)$-sparse model, which greatly reduces the number of components in (\ref{model1}). In addition, we use a rejection-control strategy within the EM algorithm to reduce the number of possible $d$-sparse combinations to $c_1 n$ where $c_1>0$ is small by Assumption \ref{aa1}. Thus instead of enumerating all $\sum_{l=1}^{d}\binom{K}{l}$ combinations, we only need to search $o(c_1 d n)$ times, which is efficient even when $K$ is large. We have a detailed discussion of the complexity of the algorithm in Section 3.1. 
\subsection{Spatial $d$-sparse Gaussian MSC}
Now let us turn our attention to some applications such as estimating the functional brain network using fMRI data and image denoising, where the input signals, $\bx_i$s, have significant spatial correlations. For this type of application, we need to incorporate the spatial correlations into the model (\ref{model1}). With a little abuse of notation, we let $\bx_i=\{x_{i1}(s_1),\dots, x_{im}(s_m)\}$ denote observations measured at $s_\ell, \ell=1, \cdots, m$, where $s_\ell\in\bR^{p}$. We then assume that 
\begin{multline}\label{model2}
\bx_i\sim \pi_1\cN(\bx_i|D(\balpha_{i1}\circ\bgamma_1), \Sigma_i)+\cdots 
+\pi_J\cN(\bx_i|D(\balpha_{iJ}\circ\bgamma_J), \Sigma_i),
\end{multline}
where $\Sigma_i$ is the covariance matrix with the ${\ell\ell'}$th entry measuring the covariance between $x_{i\ell}(s_\ell)$ and $x_{i\ell'}(s_{\ell'})$. When $p=1$, we use $\Sigma_i$ to model the temporal covariance. When $p=2$, we use $\Sigma_i$ to model the spatial covariance. In general, we assume that the covariance between random variables at two time points or two locations depends on the time lag or their inter-location distance $\Delta_{\ell\ell'}$. The most popular temporal covariance is the auto-correlated covariance which assumes that  $\{\Sigma_i\}_{\ell\ell'}=\sigma_i^2\omega_i^{-\Delta_{\ell\ell'}}$. For spatial covariance, a few popular models include the exponential model which assumes that $\{\Sigma_i\}_{\ell\ell'}=\sigma_i^2\exp{(-\omega_i\Delta_{\ell\ell'})}$ and the Gaussian model which assumes $\{\Sigma_i\}_{\ell\ell'}=\sigma_i^2\exp{(-\omega_i\Delta^2_{\ell\ell'})}$.  

\def \E {\text{E}}
\def \V {\text{Var}}
\subsection{Exponential family $d$-sparse MSC}
In general, we assume that $f(\bx_i|\theta_{ij})$ is a density in exponential family, i.e., 
\begin{equation}\label{eq:exp}
\bx_i|z_{ij}=1\sim h(\bx_i, \phi_i)\exp\{\frac{\bfeta_{ij}^\prime \bx_i-A(\bfeta_{ij})}{c(\phi_i)}\},
\end{equation}
where $\bfeta_{ij}$ is a function of the mean, $c(\phi_i)$ is the dispersion parameter and $A(\bfeta_{ij})$ is the cumulant function. For example, Gaussian distribution, binomial distribution and Poisson distribution all belong to the exponential family. When $\bx_i$ has an independent normal distribution, $\bfeta_{ij}$ is the mean of $\bx_i$ and $c(\phi_i)=\sigma_i^2$. For Poisson
and binomial models without over-dispersion, we have $c(\phi_i) = 1$. When $\bx_i|z_{ij}=1$ follows a Poisson distribution, we let $\bfeta_{ij}$ be the logarithm of its mean; when $\bx_i|z_{ij}=1$ follows a binomial distribution, we let $\bfeta_{ij}$ be the logit function of its mean.  %

To achieve $d$-sparse sparse coding, we further assume that $\bfeta_{ij}$ has the following decomposition, i.e., 
$
\bfeta_{ij} = D(\balpha_{ij}\circ\bgamma_j)
$
for a given $\bgamma_j$. 
This family of distributions has broad applications in many scientific studies, such as the RNA-seq analysis where $\bx_i$ is a binary vector on a discrete domain or network deconvolution. It can significantly broaden the application of the sparse coding algorithm. For example, we used the exponential family $d$-sparse MSC to find cancer tissue-related genetic signatures, as illustrated in Section 4.

\def \X {\textbf{X}}
\section{Fast EM Algorithm for Sparse Coding}

This section proposes a fast EM algorithm to reduce the computational cost to a linear order of $d$. 
Let $\theta_{ij}$ denote the collection of all parameters for the $i$th observation and the $j$th component. In a simple $d$-sparse Gaussian MSC, $\theta_{ij}=\{D, \balpha_{ij}, \sigma_i\}$ and in an exponential family $d$-sparse MSC, $\theta_{ij}=\{D, \balpha_{ij}, \phi_i\}$. In spatial $d$-sparse Gaussian MSC, we usually assume $\Sigma_i=\sigma_i^2R(\omega_i)$, where $R(\omega_i)$ is the correlation matrix related to spatial or temporal correlation structures. Correspondingly, $\theta_{ij}=\{D, \balpha_{ij}, \sigma_i, \omega_i\}$. Observing $\X=(\bx_1, \cdots, \bx_n)$, the likelihood function is 
$
L(\theta_{11}, \cdots, \theta_{nJ}| \X) = \prod_{i=1}^{n} \sum_{j=1}^{J}\pi_j f(\bx_i | \theta_{ij}). 
\label{eq:dsparse}
$
Let $Z$ be a $n\times J$ model-labeling matrix with the $ij$th entry $z_{ij}$. 
Then, the complete likelihood function is
$\label{eq:completelikelihood}
L(\theta_{11}, \cdots, \theta_{nJ}| \X, Z) = \prod_{i=1}^{n} \prod_{j=1}^{J} \big(\pi_j f(\bx_i |\theta_{ij})\big)^{z_{ij}}, 
$
and the log-likelihood of the complete data is 
\begin{equation}\label{loglike}
\ell(\theta| X,Z) = \sum_{i=1}^n\sum_{j=1}^J z_{ij}\big(\log\pi_j
+ \log f(\bx_i |\theta_{ij})\big).
\end{equation}

The EM algorithm is one of the most common tools for estimation in mixture models. There are a large amount of EM variants that have been proposed to facilitate the computation. A few examples include the rejection-control EM \cite{Ma08penalizedclustering}, stochastic EM \cite{nielsenEM,celeux1988random} and classification EM \cite{Celeux:1992:CEA:146597.146608}. The classical EM algorithm has two steps: E-step, which  computes the expectation of the complete-data log-likelihood function (\ref{loglike}) based on the parameters estimated in the $t$th iteration, i.e., 
$
Q(\theta|\theta^{(t)}) = \mathbb{E}_{Z|\X,\theta^{(t)}} \ell(\theta|Z, \X)$;
and M-step, in which we found $\theta^{(t+1)}$ by maximizing the function $Q(\theta|\theta^{(t)})$. 
In terms of the sparse coding, the E-step is the computation of 
\begin{equation}\label{e-step}
w_{ij}
=  \frac{\pi^{(t)}_{j}f(\bx_i|\theta_{ij}^{(t)})} {\sum_{j=1}^J \pi^{(t)}_j f(\bx_i|\theta_{ij}^{(t)})}, 
\end{equation}
and the M-step involves the maximization of 
\begin{eqnarray}
Q(\theta|\theta^{(t)}) = \sum_{i=1}^n\sum_{j=1}^J w_{ij}\log\pi_j
+\sum_{i=1}^n\sum_{j=1}^J w_{ij}\log f(\bx_i |\theta_{ij}), \label{mstep1}
\end{eqnarray}
with respect to $\theta_{ij}$, where $Q(\theta|\theta^{(t)})$ is generated by replacing $z_{ij}$ in (\ref{loglike}) by the $w_{ij}$ obtained from E-step.  
When $d$ is large, the computational cost for the conventional EM algorithm is too high. In this paper, we propose a fast EM algorithm that  updates the latent components sequentially with $d$ increasing and develop a rejection-control strategy to reduce the number of latent components for each fixed $d$. With a little abuse of notation, we use $\bgamma^{(d)}_j$ to denote the binary vector for the $j$th component under the sparsity level $d$. We first summarize the algorithm below. 

\begin{algorithm}[H]
1. Set $d=1$. Initialize the dictionary $D$ and $\pi_j=1/K$.\\
2. If $d=1$, set the $\bgamma^{(0)}_j$ as $\boldsymbol{e}_j$ where $\boldsymbol{e}_{j}$ is the unit vector with $j$th entry equal to $1$ and $S(1)$ as $\{\boldsymbol{e}_j\}_{j=1}^K$. 
If $d>1$, set $\mS(d)$ as  $\{\bgamma^{(d-1)}_j + \boldsymbol{e}_{l}\circ(\mathbbm{1} -  \bgamma^{(d-1)}_j)\mid \bgamma^{(d-1)}_j\in \mS(d-1), l=1,\dots, m\}$ . \\
3. Iteratively do E-step and M-step until convergence. \\
\textbf{E-step}: Calculate the $w_{ij}$ using (\ref{e-step}), and do the rejection-control, i.e.,
\begin{equation*}
w^\ast_{ij}=\begin{cases}
 \begin{array}{l}
 w_{ij}\quad\text{  \textbf{\textit{if}} } w_{ij} >c \quad \mbox{ \textbf{\textit{else}}} \\
  \begin{array}{l}
c\quad\text{    with probability}\  w_{ij}/c \\
    0\quad\text{      with probability}\  1-w_{ij}/c , 
 \end{array}
 \end{array}
\end{cases}
\end{equation*}
for $j=1, \cdots, J$, and then calculate the conditional expectation.\\
\textbf{M-step}: First update  $\pi^{(t+1)}_j = 1/n\sum_{i=1}^n\omega^\ast_{ij}$. Then update the rest of parameters as $\{D^{(t+1)},\balpha^{(t+1)}_{ij}, \sigma^{(t)}_i\}$ for simple Gaussian MSC, $\{D^{(t+1)},\balpha^{(t+1)}_{ij}, \sigma^{(t+1)}_i, \omega^{(t+1)}_i\}$ for spatial Gaussian MSC,  and $\{D^{(t+1)},\balpha^{(t+1)}_{ij}, \phi^{(t+1)}_i\}$ for exponential family MSC.\\
4. Update the set of the component as $\mS(d) = \{\bgamma^{(d)}_{\argmax_{j} w_{ij}}\}_{i=1}^n$.  \\ 
5. Calculated $BIC(d)$. We stop the algorithm if $BIC(d)>BIC(d-1)$, otherwise we increase $d$ by one and go to step 2.
 \caption{Fast EM algorithm for MSC}
\end{algorithm}

\subsection{Complexity of Algorithm 1}
Note that when $c = 0$, the proposed algorithm
is exactly the original EM algorithm, whereas the proposed algorithm reduces to a
variant of Monte Carlo EM algorithm \cite{wei1990monte} when $c = 1$. In practice, we first set the threshold $c$ close to $1$ at an early stage of the iterations, and then we gradually lower $c$ so that the algorithm can better approximate the original EM.

Also, we do not search all possible $d$-sparse combinations from $\sum_{l=1}^{d}\binom{K}{l}$ choices. In our algorithm, we are able to measure the number of possible combinations by $|\mathcal{S}(d)|$ where 
$\mathcal{S}(d) = \{\bm{\gamma}^{(d)}_{\argmax_{j} w_{ij}}\}_{i=1}^n$ with the nonzero entries in $\bm{\gamma}^{(d)}_{\argmax_{j} w_{ij}}$ as the dictionary vectors used to represent $\bx_i$.  $|\mathcal{S}(d)|$ is an empirical measure of $|\{\gamma_j\}|$ in Assumption \ref{aa1}.
Our searching set $S(d)$ is expanded based on the optimal dictionary set $\{\gamma_j^{(d-1)}\}$ outputted from the $(d-1)$-sparse model.
For fixed sparsity $d-1$, the cardinality of the set $\{\gamma_j^{(d-1)}\}$ is $o(n)$ by Assumption \ref{aa1}. We update the set for sparsity $d$ as $\mathcal{S}(d) = \{\bgamma^{(d-1)}_j + \boldsymbol{e}_{l}\circ(\mathbbm{1} -  \bgamma^{(d-1)}_j)\mid \bgamma^{(d-1)}_j\in \mS(d-1), l=1,\dots, m\} $ which has the cardinality less than $|\{\gamma_j^{(d-1)}\}|(K+1) =o(n(K+1))$. 
Even for $o(n(K+1))$ components, many of which have extremely small probabilities in practice. The small $w_{ij}$ can make the optimization in M-step inefficient, unstable, and sometimes even infeasible. To reduce the computational cost and stabilize the algorithm,  we incorporate a rejection-control  \cite{Ma08penalizedclustering} to shrink the small $w_{ij}$ to zero with probability $1-w_{ij}/c$. 
By Assumption \ref{aa1}, most of $n$ signals likely share same combination of atoms, which results in the reduction of the size of $\mS(d)$ to $c_1 n$ after a few iterations where $c_1$ is usually less $0.1$ in practice. The computational complexity approximates to $o( d n)$ in total.

Next, we discuss the parameter update in M-step and the information criteria in selecting $d$ for Gaussian and exponential cases separately. 

\subsection{Update dictionary for Gaussian distribution}
For (simple or spatial ) $d$-sparse Gaussian MSC model, we rewrite (\ref{loglike}) as 
\begin{multline}\label{gmstep} 
 -\sum_{i=1}^n\sum_{j=1}^J\frac{w_{ij}}{2}\big((\bx_i-D(\balpha_{ij}\circ\bgamma_j))^\prime\Sigma^{-1}_i(\bx_i-D(\balpha_{ij}\circ\bgamma_j))%
+\frac{m\log 2\pi+\log|\Sigma_i|}{2}\big). 
\end{multline}
\textbf{Update the variance:} For spatial $d$-sparse Gaussian MSC model, we have $\Sigma_i=\sigma_i^2R(\omega_i)$. 
Let $\bfeta^{(t)}_{ij}=D^{(t)}(\balpha^{(t)}_{ij}\circ\bgamma_j)$, where $D^{(t)}$ and $\balpha^{(t)}_{ij}$ are the current estimate of $D$ and $\balpha_{ij}$ respectively. Given $D^{(t)}$, $\balpha^{(t)}_{ij}$ and $\omega_i^{(t)}$, maximizing (\ref{gmstep}) with respect to $\sigma^2_i$ leads to an updated estimate of $\sigma_i^2$, which is 
${\sigma_i^2}^{(t+1)} = 1/m\sum_{j=1}^J w_{ij}(\bx_i - \bfeta^{(t)}_{ij})^{\prime} R^{-1}(\omega^{(t)}_i) (\bx_i - \bfeta^{(t)}_{ij}),
$
and for simple $d$-sparse model where $R(\omega_i)=I$, we can update $\sigma^2_i$ by 
$
{\sigma_i^2}^{(t+1)} =1/m\sum_{j=1}^{J}w_{ij}(\bx_i - \bfeta^{(t)}_{ij})^{\prime}(\bx_i - \bfeta^{(t)}_{ij}).
$\\
\textbf{Update dictionary and its coefficients:} 
Let $D_j$ be a submatrix of $D$, which collects the columns of $D$ corresponding to nonzero entries of $\bgamma_j$.
Given $\Sigma^{(t)}_i$, maximizing (\ref{gmstep}) is equivalent to solving a weighted least square regression which leads to an estimate of $\balpha_{ij}$, i.e., 
\begin{equation*}
\balpha^{(t+1)}_{ij} =\big({D_j^{(t)}}^\prime \Omega^{(t)}_iD_j^{(t)}\big)^{-1}{D_j^{(t)}}^\prime\Omega^{(t)}_i\bx_i,
\end{equation*}
where $\Omega_i^{(t)}$ is the inverse matrix of $\Sigma^{(t)}_i$ and is referred to as the precision matrix. 
For simple sparse coding model, $\Omega^{(t)}_i=\frac{1}{{\sigma_i^2}^{(t)}}I$ and $\balpha^{(t+1)}_{ij}$ has a form of a conventional least square estimate, i.e., 
\begin{equation*}
\balpha^{(t+1)}_{ij} =\big({D_j^{(t)}}^\prime D_j^{(t)}\big)^{-1}{D_j^{(t)}}^\prime\bx_i.
\end{equation*}

Notice that $|\bgamma_j|_1\le d$, which implies that $\balpha^{(t+1)}_{ij}$ is at most of dimension $d$ but not dimension $K$ as $\balpha_{ij}$. Thus, we need to transform $\balpha^{(t+1)}_{ij}$ to $K$-dimensional vector to generate final estimate of $\balpha_{ij}$ denoted by $\balpha_{ij}^{*(t+1)}$. We fill in entries of $\balpha_{ij}^{*(t+1)}$ by zero if the corresponding entries of $\bgamma_{j}$ are zero. To ease the description, we still use $\balpha^{(t+1)}_{ij}$ to denote the $\balpha_{ij}^{*(t+1)}$ in the subsequent updates. 

Next, we sequentially update each column of $D$ for given $\balpha^{(t+1)}_{ij}$  and $\Sigma^{(t)}_i$ by using a block coordinate descent algorithm. Let $\bd_k$ denote the $k$th column of the dictionary matrix $D$. Let $c_{ijk}=[\balpha^{(t+1)}_{ij}\circ\bgamma_j]_k$ where $[\cdot]_k$ is the operator to extract the $k$th entry of a vector. Now given $\mathbf{d}_k^{(t)}$, $\balpha_{ij}^{(t+1)}$ and $\Sigma_i^{(t)}$, we can update $\bd_k$ by 
\begin{equation*}
\bd_{k}^{(t+1)} =M^{-1}\sum_{i=1}^{n}\sum_{j=1}^J M^*_{ijk} \big(x_i 
-c_{ijk}\bd_{-k}\big),
\end{equation*}
where $M={\sum_{i=1}^n\sum_{j=1}^J w_{ij}c_{ijk}^2}\Omega_i^{(t)}$, $M^{*}_{ijk}=w_{ij}c_{ijk}\Omega_i^{(t)}$ and $\bd_{-k}=\sum_{l< k}\bd^{(t+1)}_l+\sum_{l> k}\bd^{(t)}_l$. Notice that $\Omega_i^{(t)}=R^{-1}(\omega_i^{(t)})/\sigma_i^{(t)}$, where $R(\omega_i^{(t)})$ quantifies the spatial correlations. 

For the simple sparse coding model where $R(\omega_i^{(t)})=I$, we can update $\bd_k$ by the simple form 
\begin{equation*}
\bd_{k}^{(t+1)} =\sum_{i=1}^{n}\sum_{j=1}^J 
\nu_{ijk}\big(x_i 
-c_{ijk}\bd_{-k}\big),
\end{equation*}
where $\nu_{ijk}=\frac{w_{ij}c_{ijk}}{{\sum_{i=1}^n\sum_{j=1}^J w_{ij}c_{ijk}^2}}$.

\textbf{Update the spatial correlation parameter:}
The final update in M-step is to maximize (\ref{gmstep}) with respect to the spatial hyper-parameter $\omega_i$ based on the updated $D$, $\balpha_{ij}$ and $\sigma^2$. Newton-Raphson is the most popular algorithm for this type of minimization. Given $D^{(t+1)}$, $\alpha^{(t+1)}$ and $\sigma_i^{(t+1)}$, we can recursively update $\omega_{i}$ by  
\begin{equation*}
\omega_{i}^{(t+1)}=  \omega_{i}^{(t)} + b^{-1}f , 
\end{equation*}
where $b= \partial^2 Q(\theta|\theta^{(t)})/ \partial^2 \omega_i$ and $f=\partial Q(\theta|\theta^{(t)})/ \partial \omega_i$.

\def \E {\text{E}}
\def \V {\text{Var}}
\subsection{Update Dictionary for distribution in exponential family}
\textbf{Update $\balpha_{ij}$:} In general, if $\bx_i|z_{ij}=1$ follows a distribution in exponential family (\ref{eq:exp}) with known dispersion parameter $c(\phi)$, the second term in (\ref{mstep1}) can be rewritten as 
\begin{equation}\label{eq:exp1}
\sum_{i=1}^{n}\sum_{j=1}^J \frac{w_{ij}}{c(\phi_i)}\big(\bfeta_{ij}^\prime \bx_i-A(\bfeta_{ij})\big),
\end{equation}
where $\bfeta_{ij}=D\balpha_{ij}\circ\bgamma_j$. Using the chain rule, the maximizer of (\ref{eq:exp1}) with respect to $\balpha_{ij}$ has the form
\begin{equation*}\label{eq:updatebeta}
\balpha_{ij}^{(t+1)} = (D^{(t)\prime}_{j} W^{(t)} D^{(t)}_{j})^{-1} D^{(t)\prime}_{j}W^{(t)} \bx_i^\ast,
\end{equation*}
where $W^{(t)}$ is an $m\times m$ diagonal matrix with $l$th diagonal entry
$$
W^{(t)}_{ll}= 1/c(\phi)\big(\partial^2 A(\bfeta^{(t)}_{ij})/\partial^2 [\bfeta_{ij}]_l\big) \big(\partial g([\bfeta^{(t)}_{ij}]_l)/\partial [\bfeta_{ij}]_l\big)^2, 
$$
where $g$ is the inverse link function, and $\bx_i^\ast= \bfeta^{(t)}_{ij} + B^{-1} (\bx_i-g(\bfeta^{(t)}_{ij}))$ of which $B$ is an $m\times m$ diagonal matrix with the $l$th diagonal entry as $\partial g([\bfeta^{(t)}_{ij}]_l)/\partial [\bfeta_{ij}]_l$. For Poisson distribution, the inverse link function is $g=\exp(\cdot)$. For binomial distribution, the inverse link function is $g(\cdot)=\exp(\cdot)/(1+\exp(\cdot))$.

\textbf{Update D:}
For general distributions of exponential family, the explicit form of $D$ is hard to obtain since the inverse link function is nonlinear. In practice, we use the gradient ascent algorithm to update $\bd_k$ by
\begin{equation*}
\bd_k^{(t+1)} = \bd_k^{(t)} + \tau U_{\bd_k},
\end{equation*}
where $U_{\bd_k}$ is the score function with respect to $\bd_k$, and $\tau$ is the step size. We use  Barzilai-Borwein method to choose a proper $\tau$ as 
$
\tau = (\bd_k^{(t)} - \bd_k^{(t-1)})^T (U_{\bd_k^{(t)}} - U_{\bd_k^{(t-1)}})/||U_{\bd_k^{(t)}} - U_{\bd_k^{(t-1)}}||^2.
$
Since our algorithm uses the value of $\bd_k^{(t)}$ in computing $\bd^{(t+1)}_k$, a single iteration has empirically been found to be enough. 
As in Gaussian sparse coding, we do not need to evaluate all the possible $d$-sparse combinations to estimate $\bd_k^{(t)}$. We only need to focus on the combinations containing the $k$th atom.

\subsection{Information criteria for selecting $d$}
The sparsity of the traditional sparse coding method is controlled by the constraint $||\balpha_i||_1 \leq \rho$ where $\rho\in(0,\infty)$. A commonly used method is to set grid points in some bounded interval $(0,c)$ and search the optimal estimation by cross validation. This approach has two difficulties: (1) the number of grid points is large; and (2) cross validation is computationally infeasible for the large data set. These difficulties can be alleviated by the MSC approach, where we use Bayesian information criterion (BIC) \cite{schwarz1978estimating} for model comparison. The minimizer of BIC score can well balance the model complexity and goodness-of-fit. In this article, we propose to use 
\begin{equation}\label{bic}
BIC(d) = -2 \ell(\hat{\theta}|X) + q(d)\log(n\times m) , 
\end{equation}
where $\hat{\theta}$ is the final estimated parameters and $q(d)$ is the number of parameters to be estimated in the model.
We set $q_{si}(d) = mK + 2n + n \sum_{j=1}^{|\mathcal{S}(d)|}\sum_{l=1}^K \bgamma_{jl}$, $q_{sp}(d) = mK + 3n +n \sum_{j=1}^{|\mathcal{S}(d)|}\sum_{l=1}^K \bgamma_{jl}$ and $q_{ex}(d)= mK + n + n \sum_{j=1}^{|\mathcal{S}(d)|}\sum_{l=1}^K \bgamma_{jl}$ as the number of parameters for simple, spatial and exponential family $d$-sparse models respectively.

\subsection{Convergence Analysis}
We consider the convergence of the proposed fast EM algorithm. Without loss of generality, we consider the simple $d$-sparse Gaussian MSC setting where $\theta=\{D,\balpha_{ij}, \sigma_i\}_{i=1,j=1}^{n,K}$ is the set of parameters.
The main difference of the $d$-sparse EM algorithm lies in the M-step where we update $D$ and $\{\balpha_{ij}, \sigma_i\}$ alternatively. In the following theorem, we show that the MSC algorithm also converges as the traditional EM algorithm. We prove the convergence of our proposed fast EM algorithm based on the results in \cite{wu1983convergence,dempster1977maximum}.

\begin{theorem}\label{thm1}
(Convergence analysis). Assume that $Q(\theta|\theta^{(t)})$ is continuous in both $\theta$ and $\theta^{(t)}$ where $\theta^{(t)}$ is an instance at the $t$th iteration. Then all the limiting points of $\{\theta^{(t)}\}_{t=0}^\infty$ are stationary points, denoted as $\theta^\ast$, and we have $L(\theta^{(t)})$ converges monotonically to $L(\theta^\ast)$ as $t$ goes to infinity for some stationary point $\theta^\ast$.
\end{theorem}
\textit{Sketch of proof}: We are aimed to show the monotonicity of the likelihood function in the iterations, that is, 
\begin{align*}
& \log\big(L(\theta; X)) - \log(L(\theta^{(t)}; X)\big)  \\
= & \big(Q(\theta; \theta^{(t)}) - Q(\theta^{(t)}; \theta^{(t)})\big) 
- \big(H(\theta; \theta^{(t)}) - H(\theta^{(t)}; \theta^{(t)})\big) , 
\end{align*}
where $H(\theta; \theta^{(t)})= E_{(\theta^{(t)})}\big(\log k(x|z,\theta)|y\big)$, $k(x|z,\theta)=L(\theta; X)/L(\theta; X,Z)$. By the proof of Theorem 1 in \cite{dempster1977maximum}, $H(\theta; \theta^{(t)}) - H(\theta^{(t)}; \theta^{(t)})<0$. In the M-step, we update the coefficients $\{\balpha_{ij},\sigma_i\}$ and the dictionary $D$ alternatively. First, we use the classical generalized linear model for computing the decomposition of $\bx_i$ over the dictionary. The uniqueness of $(\balpha_{ij},\sigma_i)$ and increment  of $Q$ function are guaranteed for this step. Next, the new dictionary $D$ is computed by column-wise, which ensures the nondecreasing of the $Q$ function.
Combining these two steps, we have the non-decreasing property of the $Q$ function. Thus, the likelihood function converges monotonically to some value $L^\ast$. We note that this convergence result can be generalized to all proposed models in Section 2 by plugging $\theta$ specified in each model. 

Then we consider the convergence rate of the proposed EM algorithm. In Theorem \ref{thm2}, we derive the convergence rate of the proposed EM algorithm. The proof is based on the convergence result in \cite{meng1994global}. Notably, we show that the proposed EM algorithm has a linear iteration with the rate $r$ related to the Jacobian matrix $J(\theta^\ast)$. 

\begin{theorem}\label{thm2}
(Convergence rate) Assume that $Q(\theta|\theta^{(t)})$ is continuous in both $\theta$ and $\theta^{(t)}$ where $\theta^{(t)}$ is an instance at the $t$-th iteration. Let $\theta^\ast$ be the stationary point. The convergence rate $r = \lim_{k\to \infty} || \theta^{(t+1)} - \theta^\ast||/||\theta^{(t)} - \theta^\ast||$ is 
\begin{equation*}
r = \lambda_{\max} \equiv \mbox{the largest eigenvalue of } J(\theta^{\ast}) , 
\end{equation*}
where $J(\theta^{\ast})= \{J_{ij}(\theta^\ast)\}$ with $J_{ij}(\theta^\ast) = \partial \mathbf{M}_i(\theta^\ast)/\partial\theta^\ast_{j}$ and $\mathbf{M}_i(\theta^\ast)$ is the $i$-th element of $\mathbf{M}(\theta^\ast)$ defined as $\mathbf{M}(\theta^\ast) = \arg\max_{\theta\in \Theta} Q(\theta| \theta^\ast)$.
\end{theorem}

\textit{Sketch of proof}: By Theorem \ref{thm1}, we have $\theta^{(t)}$ converges to $\theta^\ast$ and that $M(\theta)$ is differentiable at $\theta^\ast$. The Taylor expansion gives
\begin{equation*}
\theta^{(t+1)} - \theta^\ast = (\theta^{(t)} - \theta^\ast) J(\theta^{\ast}) + O(||\theta^{(t)} - \theta^\ast||^2) , 
\end{equation*}
where $J(\theta^\ast)$ is defined in Theorem \ref{thm2}. By Poroposition 1 in \cite{meng1994global}, we have 
\begin{equation*}
r =  \mbox{the largest eigenvalue of } J(\theta^{\ast}).
\end{equation*}

\section{Empirical Studies}
\subsection{Gaussian $d$-sparse model}
\textbf{Simulation: }Fifty mock data sets were generated to compare the empirical performance of our MSC method with the popular sparse coding methods such as K-SVD and online dictionary learning. We generated each signal $\boldsymbol{x}_i\in \mathbb{R}^m$, where $m=100$, from a mixture of normal distribution $\sum_{j=1}^J\pi_j\mathcal{N}(D(\balpha_{ij}\circ\bgamma_j),\sigma^2R(\omega))$, $i = 1,\ldots,n$. Each element of the dictionary matrix $D\in \mathbb{R}^{m\times K}$, where $K=30$, was fixed realization from $\text{Uniform}[0,1]$, and every column of $D$ was normalized. Each element of $\balpha_{ij}$ was generated from $\text{Uniform}[1,10]$ and was kept fixed once generated. The weight $\pi_j$ was set to $1/J$, where $J=\sum_{l=1}^2\binom{30}{l}$, for $j=1,\ldots,J$. The spatial locations of signals were randomly realized from the $[0,100]^2$ spatial domain. The exponential correlation function with $\omega=1/25$ was employed to model the spatial correlation of signals.  We set $\sigma^2=||D(\balpha_{ij}\circ\bgamma_{j})||_2/\text{SNR}$, where $\text{SNR}=2,4$ respectively.

\begin{figure*}[h!]
\centering
\includegraphics[width=\textwidth]{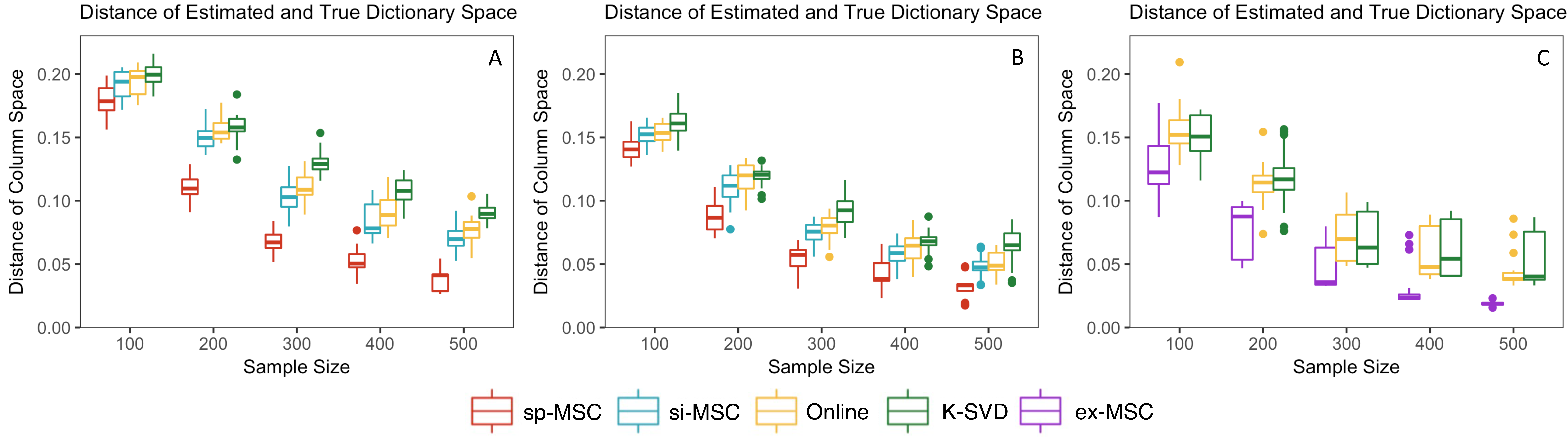}
\caption{Figure A and B: Estimation error of data with Gaussian distribution, $\text{SNR}=2, 4 $ respectively. Figure C: Estimation error of data with Poisson distribution.}\label{sim}
\end{figure*}

We implemented the spatial $d$-sparse Gaussian MSC (sp-MSC) algorithm, the simple $d$-sparse Gaussian MSC (si-MSC) algorithm, the online dictionary learning (Online) algorithm, and the K-SVD algorithm on the synthetic signals with sample size $n$ varying from $100$ to $500$. In Figure~\ref{sim}(A)-(B), we drew boxplots of the distance between the space spanned by the estimated dictionary and the space spanned by the true dictionary for different sample sizes. The sp-MSC algorithm outperformed other algorithms in terms of average distance and standard deviation. Moreover, as we increased the noise level, the sp-MSC algorithm had a significant advantage over other comparable algorithms, which implies that the sp-MSC algorithm is especially useful for noisy data. 

\begin{figure*}[h!]
\centering
\includegraphics[width=0.8\textwidth]{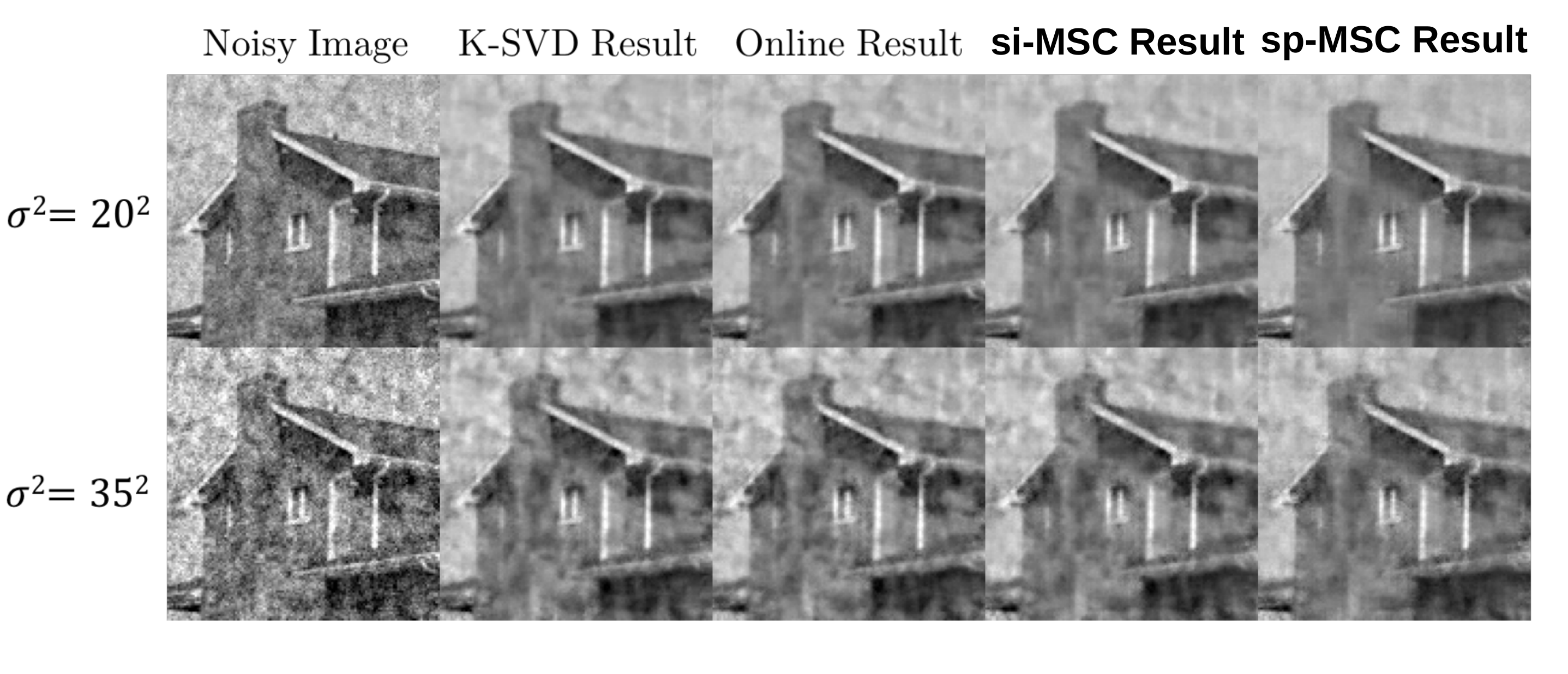}
\caption{Plotted in the columns are generated images (first column), donoised images (2-5 columns) using K-SVD,  online dictionary learning (Online), simple $d$-sparse Gaussian MSC (si-MSC) and spatial $d$-sparse Gaussian MSC (sp-MSC) respectively.}\label{image}
\end{figure*}

\begin{figure*}[h!]
\centering
\includegraphics[width=0.8\textwidth]{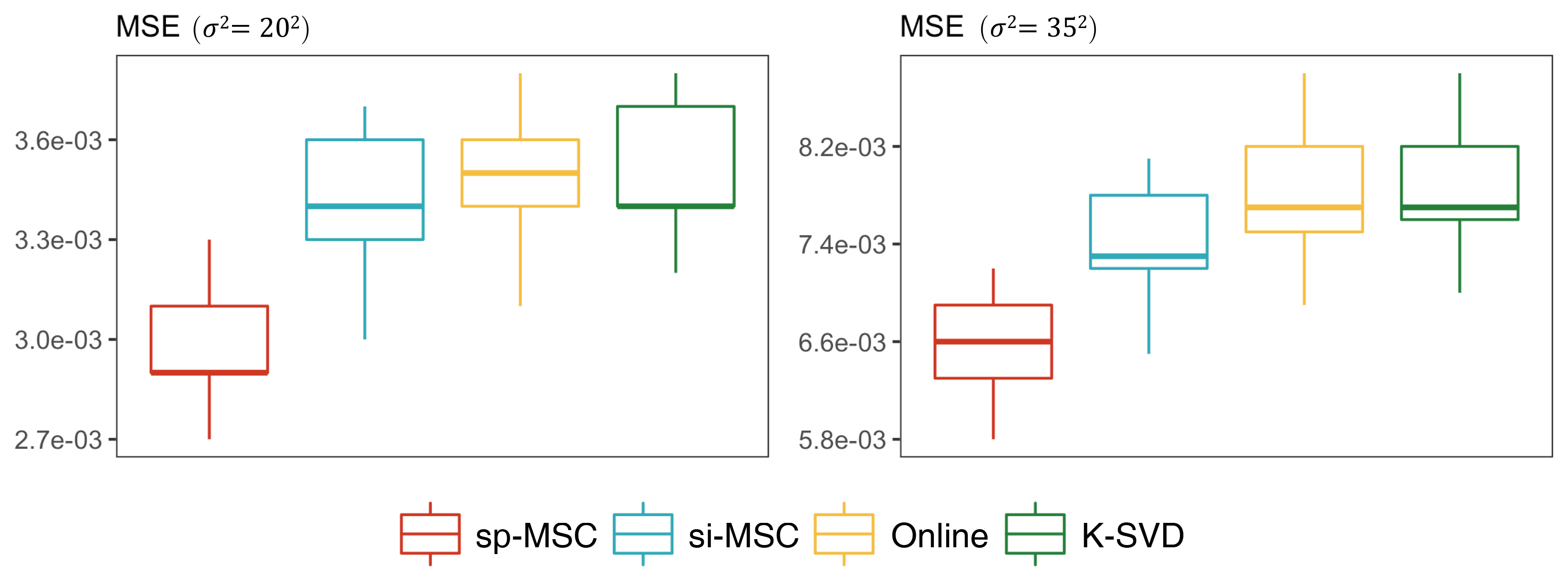}
\caption{Plotted here are the MSE of denoised images using the spatial $d$-sparse Gaussian MSC (sp-MSC) algorithm, the simple $d$-sparse Gaussian MSC (si-MSC) algorithm, the online dictionary learning (Online) algorithm and the K-SVD algorithm.}\label{mse}
\vspace{-10pt}
\end{figure*}

\textbf{Application (Image denoising):}
In this example, five $128 \mbox{pixel}\times128 \mbox{pixel}$ images were used for denoising. Noise from a Gaussian random field with covariance function $\{\Sigma_i\}_{\ell\ell'}=\sigma^2\exp{(-1/4\Delta_{\ell\ell'})}$ was artificially added to the raw images. In Figure \ref{image}, we compared the denoised images at $\sigma^2=20^2$ and $\sigma^2=35^2$. We have $n=1600$ overlapping blocks with intensity of which were stretched as a $m=144 (12\mbox{pixel}\times 12\mbox{pixel})$ dimensional vector. Clearly, each entry in the vector was spatially correlated.  We calculated the mean squared error (MSE) for sp-MSC, si-MSC, online dictionary learning, and K-SVD. Figure \ref{mse} clearly shows that sp-MSC significantly outperformed comparable algorithms in terms of estimation error. Both sp-MSC and si-MSC outperformed the existing algorithms, where the sp-MSC has a better estimation performance for spatially correlated data. 

\textbf{Application (Brain connectivity study use fMRI data):}
Understanding the organizational architecture of human brain function has been of intense interest. After decades of active research using in-vivo functional neuroimaging techniques such as fMRI, there is accumulating evidence that human brain function emerges from and is realized by the interaction of multiple concurrent neural processes or networks, each spatially distributed across the specific structural substrate of neuroanatomical areas. Although this discovery holds a lot of promise on constructing the concurrent functional networks and network-level interactions robustly and faithfully at the whole population level, the delivery of this promise, however, has not yet been fully materialized due to the lack of effective and efficient analytical tools for handling the spatially correlated brain image data. 

\begin{figure*}[h!]
\centering
\includegraphics[width=\textwidth]{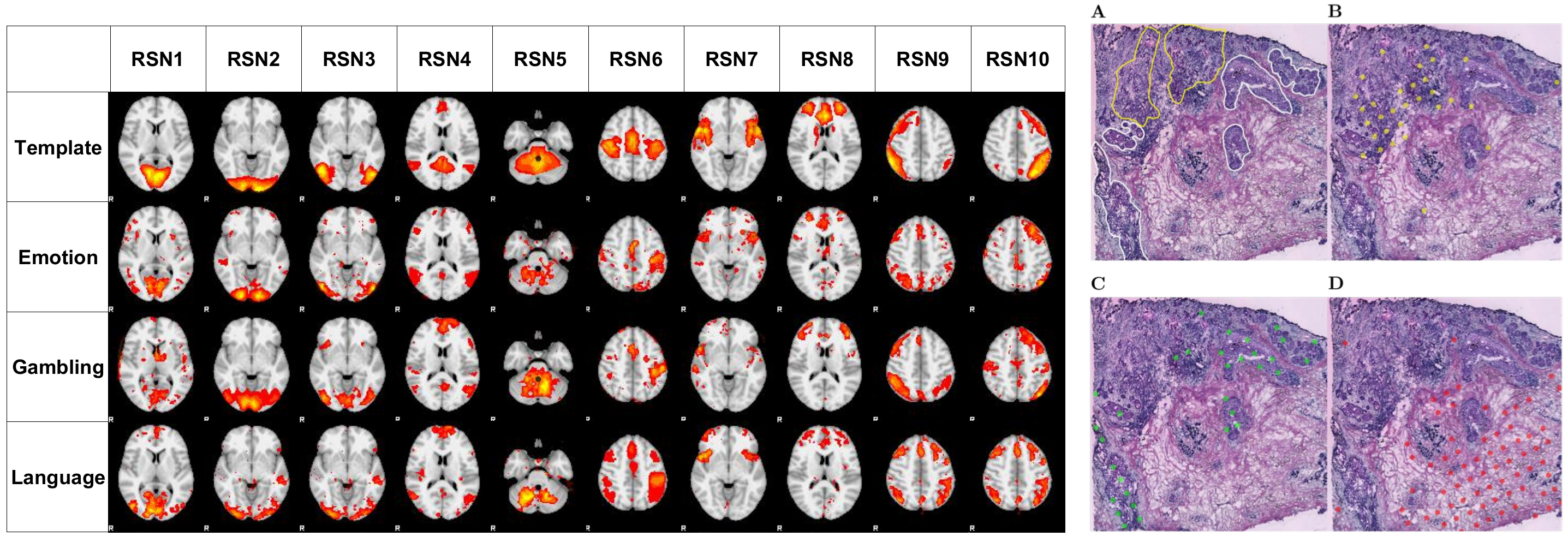}
\caption{Left: Ten Resting-state networks (RSNs 1-10) identified by the sp-MSC algorithm. Right: Plotted in A is the histological section of a breast cancer biopsy with invasive ductal cancer areas (yellow line), ductal cancer in situ areas (white line), and non-cancer areas (other areas). We plotted the predicted invasive ductal cancer areas in B, ductal cancer in situ areas in C, and non-cancer areas in D.}\label{fmri}
\end{figure*}

We applied the proposed spatial $d$-sparse Gaussian MSC on the Human Connectome Project (HCP) Q1 released functional magnetic resonance imaging (fMRI) data to meet this challenge. Three tasks (``Emotion", ``Gambling" and ``Language" ), each including fMRI images of $5$ subjects, were selected to demonstrate how MSC can help to understand the human brain connectivity. The data were preprocessed using FSL \cite{jenkinson2012fsl}. Using $K=100$ and Gaussian spatial correlation function $\{\Sigma_i\}_{\ell\ell'}=\sigma_i^2\exp{(-\omega_i\Delta^2_{\ell\ell'})}$, we found the optimal sparse level $d=5$ based on the BIC criteria defined in (\ref{bic}). 
For each task, we mapped the learned atom $\{\bd_k\}_{k=1}^{100}$ in the dictionary on the brain and compared them with the resting state networks (RSNs) \cite{smith2009correspondence,smith2012temporally}. The intrinsic RSNs have also been observed in task-based fMRI data \cite{lv2015modeling,smith2009correspondence}. As shown in the left panel of Figure \ref{fmri}, we selected $10$ learned networks corresponding to the (RSNs 1-10) using task-based fMRI data.  RSNs $1$-$3$ mainly include the visual cortex; RSN $4$ is often referred to as the default mode network; RSN $5$ covers the cerebellum; RSN $6$ dominantly features sensor-motor network; RSN $7$ covers the auditory network; RSN $8$ covers the executive control network; the symmetric RSN $9$ and RSN $10$ cover the left and right middle frontal, orbital and superior parietal areas, while for certain tasks (e.g. Language) it is observed that RSN $9$ and RSN $10$ will merge into the same network.

\subsection{Poisson $d$-sparse Model}

\textbf{Simulation:}
Fifty mock data set were simulated from a mixture of Poisson distribution\\ 
$\sum_{j=1}^J\pi_j \text{Poisson}(\theta_{ij})$, where $\log\theta_{ij}=D(\balpha_{ij}\circ\bgamma_j)$ with $D\in \mathbb{R}^{100\times 10}$. The dictionary $D$, coefficient $\balpha_{ij}$ and weight $\pi_j$ were generated in the same way as the Gaussian $d$-sparse simulation. We implemented the exponential family $d$-sparse MSC (ex-MSC) algorithm, the online dictionary learning (Online) algorithm, and the K-SVD algorithm on the synthetic signals with sample size $n$ varying from $100$ to $500$. The distance between the column space of an estimated dictionary and the column space of the true $D$ was plotted in Figure~\ref{sim}(C). Our exponential $d$-sparse MSC algorithm provided a significantly better estimate of the true dictionary.

\textbf{Application (Spatial Transcriptomic imaging for Breast Cancer Data):}
In this real data analysis, we applied the ex-MSC algorithm to a breast cancer spatial transcriptomics dataset \cite{2016Sci...353...78S}. Spatial transcriptomics is a recent sequencing strategy that quantifies the gene expression within a tissue section with two-dimensional positional information. The sequenced reads are aligned to the reference genome in the dataset to count the number of reads mapped to a specific gene. We selected $1573$ genes with reads count larger than $100$ at $254$ locations in a histological section of a breast cancer biopsy, including the invasive cancer areas, cancer in situ areas, and the non-cancer areas. The input data vectors $\bx_i\in\bR^{m}$ were the mapped reads count at $m$ locations in a histological section for $i=1,\dots,n$ where $n$ is the number of selected genes.

We applied the ex-MSC algorithm to this dataset with dictionary size $K=10$. The optimal $d$ was set as two based on the BIC criteria.  The learned atoms were mapped back to the tissue image and plotted in the right panel of Figure~\ref{fmri}. 
There is a strong association between atoms and cell types.  
Three atoms were drawn on the same histological section of breast cancer biopsy, with the first one representing the gene expression of the invasive cancer areas (Figure~\ref{fmri} B), the second one representing the gene expression of cancer in situ areas (Figure~\ref{fmri} C) and the last one representing the gene expression of the non-cancer areas (Figure~\ref{fmri} D). Comparing with traditional cell-type identification methods with human supervision, the ex-MSC provides a data-driven way for pathological analysis.

\section{Discussion}
\label{sec:diss}

The contribution of the MSC procedure is two-fold. First, it can deal with any observations: continuous observations or discrete observations. The mixture model framework that the MSC procedure relies on includes a classical sparse coding algorithm as a special case. Therefore, MSC can be considered a generalization of the classical sparse coding approach to the general data format. Second, as demonstrated by our simulation studies, MSC can effectively handle spatially correlated predictors, challenging existing sparse coding methods. If practitioners are unsure about the spatial or temporal information, we suggest beginning with the spatial or temporal $d$-sparse model. Based on the magnitude of the off-diagonal entries in the estimate covariance, we may decide to choose to include the spatial or temporal structure or not. A more rigorous way is to develop a statistical test for adaptive selection based on testing whether the covariance matrix is a diagonal matrix or not. This is out of the scope of the current paper, and we are interested in exploring the test in our future work. In addition, the proposed fast EM algorithm reduces the number of the latent component to the linear order of $d$ and only needs to estimate a finite number of sparse models instead of searching tuning parameters in a continuous domain using cross-validation.

\bigskip

\section*{References}

\bibliographystyle{apalike}
\bibliography{IEEEabrv,ref}

\end{document}